# Decoupled domain-texture switching from magnetic easy axis in kagome ferromagnet $EuTi_3Bi_4$

Yunhao Wang[1,2,#], Shiyu Zhu[1,2,#]✉, Guohao Xi[3,4,#], Runnong Zhou[1,2,#], Ruwen Wang[1,2], Jianfeng Guo[1,2], Jiali Liu[1,2], Zichao Chen[1,2], Kailin Xu[1,2], Cong Wang[3]✉, Chengmin Shen[1,2], Jiang Xiao[4], Haitao Yang[1,2], Xiaoli Dong[1,2], Wei Ji[3]✉, Hong-Jun Gao[1,2]✉

[1] Beijing National Center for Condensed Matter Physics and Institute of Physics, Chinese Academy of Sciences, Beijing 100190, China.

[2] School of Physical Sciences, University of Chinese Academy of Sciences, Beijing 100190, China.

[3] Key Laboratory of Quantum State Construction and Manipulation (Ministry of Education), School of Physics, Renmin University of China, Beijing 100872, China

[4] Department of Physics and State Key Laboratory of Surface Physics, Fudan University, Shanghai 200433, China

## Abstract

Magnetic anisotropy defines the easy axis of a magnetic material and governs the spatial arrangement of its domains. To date, anisotropy engineering has focused on reorienting the easy axis or tuning the anisotropy energy, both of which demand substantial energy input. Here, we demonstrate that magnetic domain textures can be switched without reorienting the easy axis, as observed in a kagome ferromagnet $EuTi_3Bi_4$ crystal. Using low-temperature magnetic force microscopy, we observe that the preferred orientation of magnetic domains switches from the *a*-axis to the *b*-axis upon temperature variation, and that this switching can also be triggered by an out-of-plane magnetic-field reset. Magnetization measurements and density functional theory calculations confirm a robust *c*-axis easy magnetization, ruling out a conventional spin-reorientation transition. Instead, the texture switching is governed by the temperature dependence of the in-plane variation of the Magnetic anisotropy energy landscape, which arises from two competing interactions with different decay rates: single-ion anisotropy favors *a*-oriented spin components, while nearest-neighbor anisotropic exchange favors *b*-oriented ones. Furthermore, the critical switching temperature is substantially elevated in a mechanically exfoliated $EuTi_3Bi_4$ flake. Our findings establish that macroscopic magnetic textures can be effectively manipulated by tuning the competition between in-plane anisotropic interactions, without the energy cost of reorienting the easy axis.

## I. Introduction

Magnetic anisotropy—the tendency of magnetization to align along specific crystallographic directions—is central to magnetism. [1–8]. It defines the magnetic easy axis, sets the energy scale for magnetization reversal, and governs the spatial arrangement of magnetic domains. A variety of established strategies exist to tailor magnetic anisotropy, including doping, strain, electrostatic gating, and interfacial engineering ferroelectric polarization [9,10], light [11,12], electric gating [13,14], voltage [15,16] and strain [17–20], and they have proven effective in modifying both the direction of the easy axis and the magnitude of the anisotropy energy.. These methods hold great promise for future applications in information storage and processing [21,22], leveraging spin-dependent electronic effects such as anisotropic magnetoresistance. Reorienting the easy axis, however, requires overcoming the entire energy barrier that separates distinct easy-axis directions, which demands substantial external stimuli. Tuning the anisotropy energy, on the other hand, primarily affects critical fields and transition temperatures. The anisotropy energy, however, is not merely a single number. As a function of the magnetization direction, it defines a multidimensional energy landscape whose finer features—local minima along directions other than the easy axis—are rarely examined. Modulating these subtle features requires far less energy than reorienting the dominant easy axis, yet whether such modulation can produce observable macroscopic switching has remained unclear.

Kagome magnets have emerged as a promising platform for exploring complex magnetism arising from the interplay of geometric frustration, topology, and strong correlations [23–31]. Recently, the newly discovered $R\mathrm{Ti_3Bi_4}$ ($R$: rare-earth metals) kagome family, featuring a layered structure and quasi-one-dimensional zigzag chains of $R$ atoms, has attracted considerable attention as a fertile ground for uncovering complex magnetic phenomena. Pronounced magnetic anisotropy [32–34], spin density waves [35,36], and other emergent properties [37–40] have been reported in this system. In particular, $\mathrm{EuTi_3Bi_4}$, which combines a van der Waals structure, out-of-plane ferromagnetism, and rich magnetic properties, offers an especially suitable platform for the study and manipulation of magnetic anisotropy governed by intra- and inter-chain magnetic coupling of Eu zigzag chains [41–43].

Here we report the first discovery of a new-type in-plane biaxial magnetic domain-texture switching (MDS) phenomenon, controlled by magnetic field and temperature, in $\mathrm{EuTi_3Bi_4}$ crystal using ultra-low temperature magnetic force microscopy (MFM) combined with vector magnetic field. In the $\mathrm{EuTi_3Bi_4}$ crystal, the energetically preferred magnetic domain orientation undergoes a transition from the $a$-axis to the $b$-axis at a characteristic switching temperature $T_s$. Importantly, this switching is not a reorientation of the easy axis. Magnetization measurements and density functional theory (DFT) calculations show that the easy axis remains firmly along *c*-axis, while the in-plane magnetic anisotropy reveals a competition between two distinct energy preferences. Theoretical analysis traces this competition to the single-ion anisotropy and the nearest-neighbor anisotropic exchange, which favor the *a*- and *b*-oriented spin components, respectively. As temperature changes, these two interactions decay at different rates, causing the preferred domain texture to switch between the two axes. Furthermore, $T_s$ is significantly enhanced in a mechanically exfoliated $EuTi_3Bi_4$ flake, underscoring the high tunability of magnetic anisotropy in $EuTi_3Bi_4$ and its promising potential for spintronic applications. Our report of this new form of magnetic anisotropy regulation that enables macroscopic reorientation of magnetic textures without reorienting the easy axis, introduces a novel concept of magnetic control and paves the way for the development of low-energy spintronic devices.

## II. Magnetic domain orientation switch under varying out-of-plane field in $EuTi_3Bi_4$

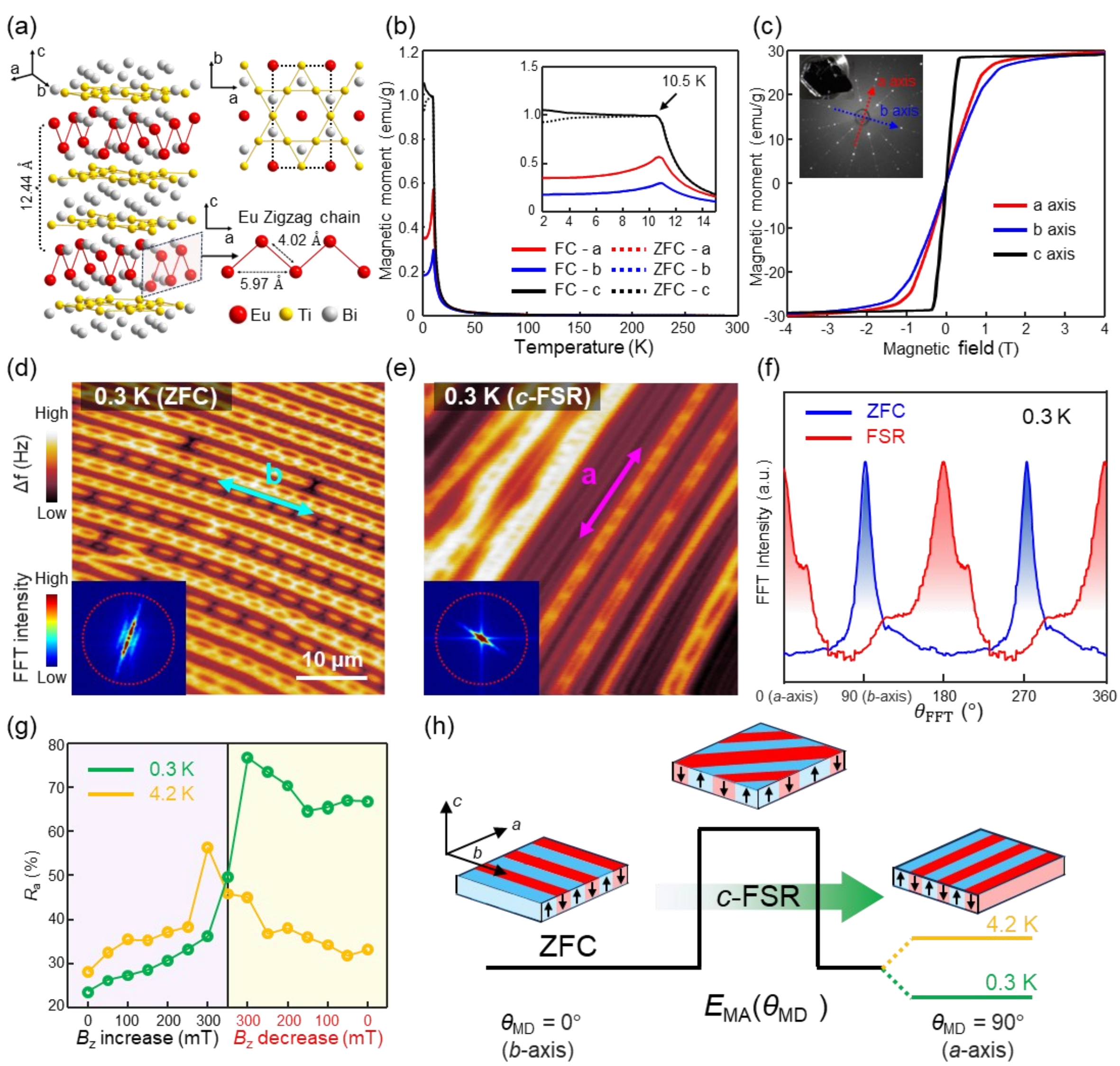

**Fig.1. Out-of-plane field induced MDS of $EuTi_3Bi_4$.** (a) Crystal structure of $EuTi_3Bi_4$ crystal. (b) Temperature-dependent magnetization curves of $EuTi_3Bi_4$ measured under an external magnetic field of 10 mT. Zero-field-cooled (ZFC) and field-cooled (FC) data with the field applied along the *a*/*b*/*c*-axis are presented. The inset highlights a zoomed-in view of the temperature range around the Curie temperature. (c) Field-dependent magnetization curves of $EuTi_3Bi_4$ at 2 K, under a magnetic field along *a*/*b*/*c*-axis of the crystal, respectively. Inset shows the *a*/*b*-axis orientations of the crystal as determined by Laue diffraction. (d) MFM images of magnetic domains in $EuTi_3Bi_4$ crystal after ZFC to 0.3 K, inset shows corresponding FFT image. (e) MFM images of magnetic domains in $EuTi_3Bi_4$ crystal after out-of-plane FSR at 0.3 K, inset shows corresponding FFT image. (f) $I_{FFT}$ curve extracted by integrating FFT intensity within red circles from FFT images in (d) and (e), revealing significant *b*-axis and *a*-axis orientations, respectively. (g) $B_z$–dependent *a*-axis ratio component $R_a$ curve at 0.3 K and 4.2 K. (h) Schematics showing magnetic anisotropy energy $E_{MA}$ as a function of magnetic textures with corresponding magnetic domain orientation $\theta_{MD}$.

$EuTi_3Bi_4$ is a member of the emergent $RTi_3Bi_4$ kagome family featuring an orthorhombic lattice composed of Ti-based kagome layers and Eu zigzag chains along *a*-axis (Fig 1a). The $EuTi_3Bi_4$ crystals studied in this work are grown via a self-flux method. Topography images of the exfoliated $EuTi_3Bi_4$ crystal measured by scanning tunneling microscopy (STM) reveal a step height of 1.40 nm and a neighboring atom distance of 0.62 nm (Fig S1). X-ray diffraction (XRD) and energy-dispersive x-ray spectroscopy (EDS) confirm high quality of the single crystal (Fig S2). Magnetization and transport measurement reveal dominant out-of-plane ferromagnetism and Curie temperature $T_c$ of the $EuTi_3Bi_4$ crystal at around 10.5 K, as well as significant magnetic anisotropy with the easy axis along the *c*-axis. (Fig 1b-c and Fig S3-S4).

The $EuTi_3Bi_4$ crystal is firstly zero-field-cooled (ZFC) to 0.3 K, below its Curie temperature. Subsequent MFM measurements reveal an ordered magnetic domain structure composed of aligned stripe domains and magnetic bubble chains (Fig 1d). The orientation of this composite magnetic structure is identified to align along the *b*-axis of the crystal, as confirmed by Laue diffraction (inset of Fig 1c). Next, an out-of-plane field saturation reset (FSR) procedure is carried out, in which the applied magnetic field is first increased to fully saturate the sample and then reduced to zero. As the out-of-plane magnetic field $\boldsymbol{B}_\mathbf{z}$ increases, the magnetic domain structure gradually fades and completely disappears at saturation field $\boldsymbol{B}_\mathbf{z}$ = 350 mT (Fig. S5a). Upon reducing the magnetic field back to 0 T, a new magnetic domain pattern preserving the characteristic stripe and bubble chain morphology re-emerges. Strikingly, the new magnetic domains are reoriented along the *a*-axis (Fig. 1e). In a consistent experiment conducted at 4.2 K, the original *b*-axis-oriented magnetic texture reappears after ZFC, as expected. Contrary to the situation at 0.3 K, the *b*-axis-oriented chain-like magnetic texture gradually recovers following the FSR process at 4.2 K (Fig. S5b). The MDS phenomenon observed at 0.3 K—where the magnetic domain orientation switches between crystallographic axes—together with the inconsistent FSR results between 0.3 K and 4.2 K, indicates a temperature-dependent anisotropic magnetic transition.

Fast Fourier transform analysis is performed to quantitatively evaluate the magnetic domain orientation switching behavior. For each MFM image representing a specific magnetic domain structure, the integrated FFT intensity $I_{\mathrm{FFT}}(\theta_{\mathrm{FFT}})$ is obtained from the corresponding FFT image by integrating the intensity from the Γ point over a suitable radial range that captures the dominant features along a given in-plane direction $\theta_{\mathrm{FFT}}$. The resulting $I_{\mathrm{FFT}}$ curves reveal a pronounced peak shift between the ZFC and FSR conditions at 0.3 K, with a peak position difference of 90° (Fig. 1f). These two peak positions correspond to magnetic domain orientations along the crystallographic *b*-axis and *a*-axis, respectively.

Further, significant difference during FSR procedures between 0.3 K and 4.2 K can be visualized from *a*-axis component $R_a(\boldsymbol{B}_z)$ curve (Fig 1g). $R_a$ is extracted from $I_{FFT}$ curves and is defined as $R_a = I_a/(I_a+I_b)$, where $I_a$ and $I_b$ represent the integrated FFT intensities within a 30° range centered around $\theta_{FFT} = 0°$ (*a*-axis) and $\theta_{FFT} = 90$ (*b*-axis), respectively. At 4.2 K, $R_a$ remains consistently low throughout the entire FSR process, indicating a persistent preference for *b*-axis domain orientation. In contrast, at 0.3 K, a distinct jump in $R_a$ is observed between the increasing and decreasing $\boldsymbol{B}_z$ stages, signifying a switch in magnetic domain orientation.

The observed *b*-axis and *a*-axis–oriented textures induced through ZFC or FSR operations, reveal a pronounced anisotropy in magnetic domain orientation. The system resides in an energy minimum when magnetic domains align along either the *a*- or *b*-axis, while the application of an out-of-plane magnetic field can facilitate transitions between these energetically stable states. More intriguingly, the comparable energy between *a*-axis and *b*-axis oriented magnetic textures is effectively tuned by temperature. (Fig. h).

## III. Temperature-dependent domain orientation switch

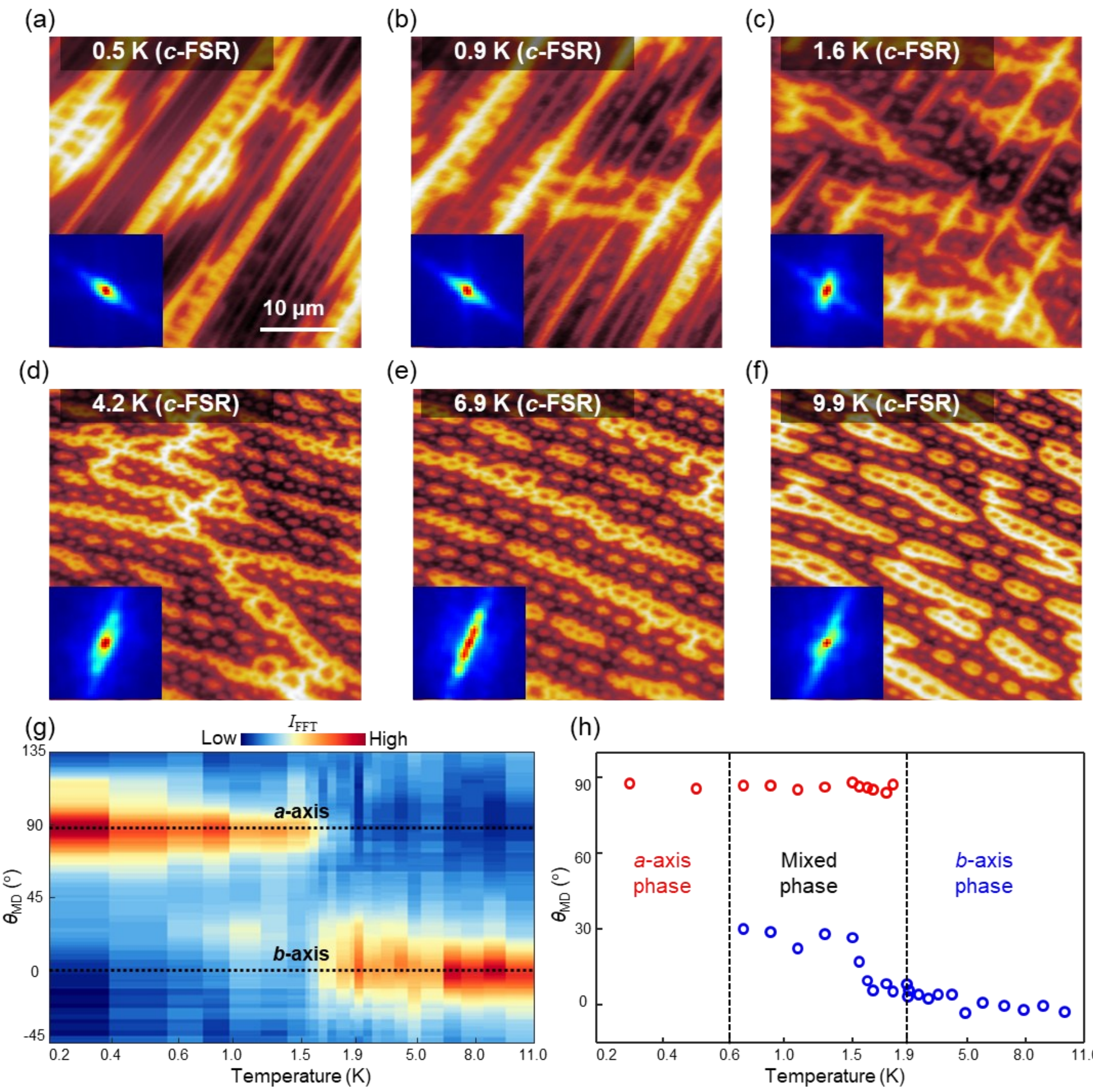

**Fig.2. Temperature-dependent domain orientation switch of $EuTi_3Bi_4$ crystal.** (a)-(f) MFM images of magnetic domains after out-of-plane FSR at different temperatures. (g) Temperature-dependent magnetic phase diagram revealing significant magnetic domain orientation switch at around 1.6 K. (h) Extracted magnetic domain orientation component from (g).

In-depth temperature-dependent *c*-axis FSR experiments reveal a detailed temperature-dependent magnetic domain orientation variation. At low temperature, FSR-induced magnetic domains preferentially align along *a*-axis, forming typical stripe domain structures (Fig 1e and Fig 2a-b). As temperature increases, *b*-axis oriented branch-like magnetic domains gradually increases. Near 1.6 K, both *a*- and *b*-axis domain orientations coexist with comparable weight (Fig 2c). At higher temperatures, magnetic domains predominantly align along the *b*-axis (Fig 2d-f).

Clear MDS behavior at a switching temperature of $T_s = 1.6$ K can be visualized in a temperature-dependent magnetic phase diagram (Fig 2g), constructed by a series of $I_{FFT}$ curves extracted from MFM images (Fig S6) acquired after *c*-axis FSR from 0.3 K to 10.0 K. By extracting fitted peak position, the phase diagram can be further divided into three regions: *a*-axis phase below around 0.6 K, where magnetic domain tend to align along *a*-axis; *b*-axis phase between 1.9 K and 10.0 K, where magnetic domains tend to align along *b*-axis; and a mixed phase between 0.6 K and 1.9 K, where components of *a*-axis and *b*-axis magnetic domain orientation coexist (Fig 2h). In the mixed phase, the component associated with the *b*-axis orientation is significantly weaker than the *a*-axis component, and its orientation clearly deviates from the exact *b*-axis direction. Complete MFM images of temperature-dependent out-of-plane FSR procedures are listed in Fig S7.

The phase diagram further explains the MDS phenomenon induced by *c*-axis FSR at $T < T_s$. Magnetic domains aligned along the *a*-axis and *b*-axis represent the lowest-energy configurations below and above $T_s = 1.6$ K respectively. During a ZFC process, the $EuTi_3Bi_4$ crystal initially passes through the temperature regime $T_s < T < T_c$, where magnetic domains preferentially align along the *b*-axis. Upon further cooling below $T < T_s$, the *a*-axis becomes the energetically favorable direction for magnetic domain alignment. However, the preexisting *b*-axis-oriented domains remain as a metastable state. A subsequent *c*-axis FSR process assists the system in overcoming the energy barrier within the magnetic domain orientation range 0° (*b*-axis) < $\theta_{MD}$ < 90° (*a*-axis), facilitating the formation of *a*-axis-oriented magnetic textures.

## IV. Mechanism of magnetic domain anisotropy

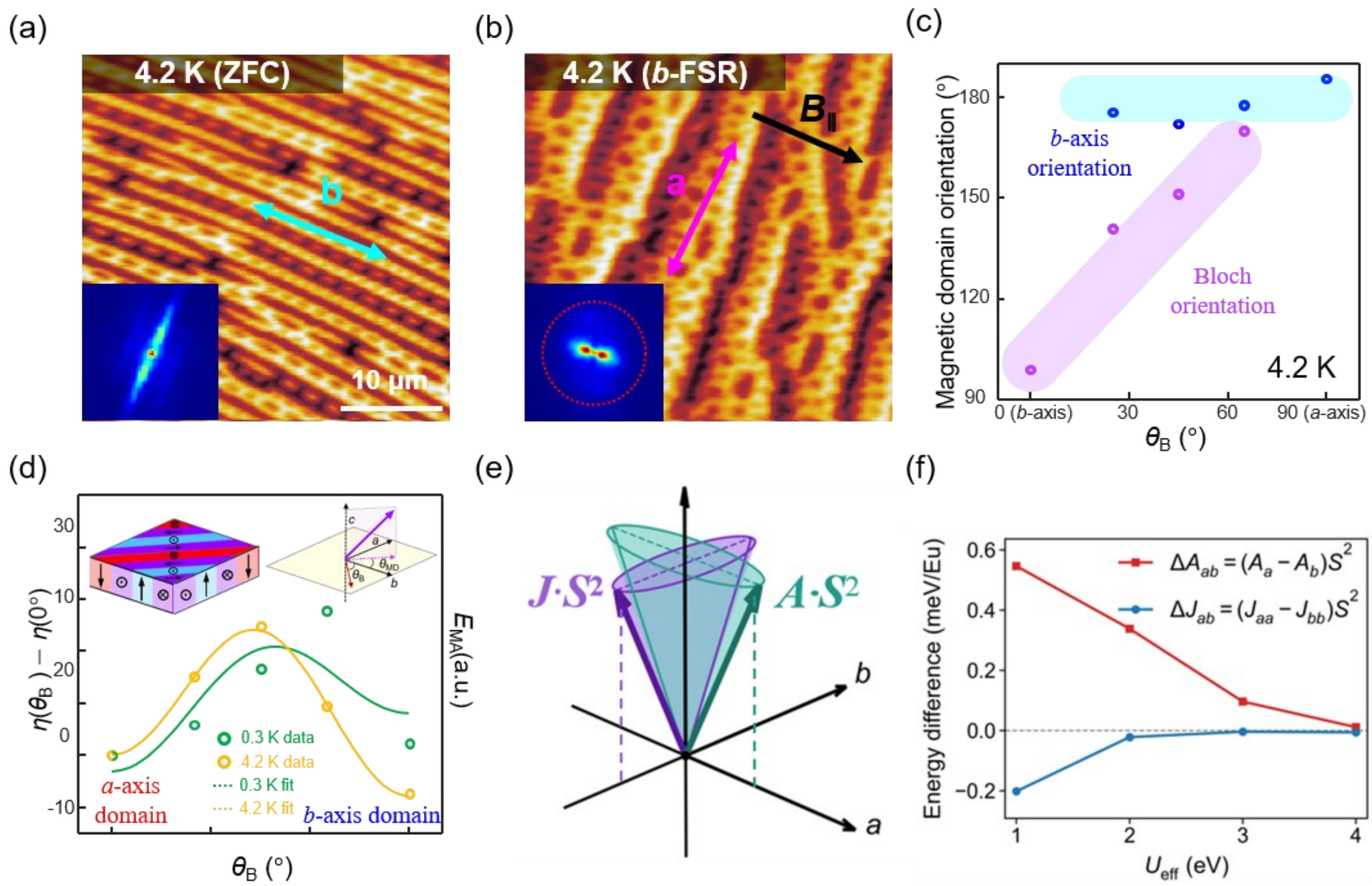

**Fig.3. In-plane anisotropy model.** (a) Magnetic domain textures in $EuTi_3Bi_4$ crystal after ZFC to 4.2 K. (b) Magnetic domains formed after FSR along *b*-axis at 4.2 K. Insets of (a)-(b) show corresponding FFT images. (c) Extracted magnetic domain orientation component from FSR images under magnetic field along different $\theta_B$. Blue and purple stripes represent *b*-axis orientation and orientation perpendicular to $\theta_B$ (Bloch-type domain wall), respectively. (d) The extracted field regulation factor $\eta$ following in-plane FSR results along variable directions at 0.3 K and 4.2 K, dashed curves indicate the orientation-dependent variation of $E_{MA}$ within the plane, the inset illustrates the magnetic texture model, the purple arrow represents magnetic moment in the domain wall. (e) Schematic of finite-temperature canting of the Eu moments around the *c*-axis easy direction. The green and purple ellipses depict the in-plane canting angle preferences produced by the single-ion anisotropy *A* and the nearest-neighbor anisotropic exchange *J*, respectively. (f) $U_{eff}$ dependence of $\Delta J_{ab}=(J_{aa}-J_{bb})S^2$ and $\Delta A_{ab}=(A_a-A_b)S^2$.

To further investigate the underlying mechanism of MDS phenomenon, in-plane FSR operations are performed. At 4.2 K, in-plane magnetic field $\boldsymbol{B}_{\parallel}$ with direction $\theta_B$ = 0° (*b*-axis) is applied on a $EuTi_3Bi_4$ crystal initially exhibiting ZFC-induced *b*-axis–oriented magnetic domains. After increasing $\boldsymbol{B}_{\parallel}$ to fully saturate the magnetization and subsequently reducing it to 0 T, the re-emerged magnetic domains are switched to *a*-axis orientation (Fig 3a-b). Additional in-plane FSR procedures are carried out at more in-plane applied field directions: $\theta_B$ = 25°, 45°, 65°, 90°(*a*-axis), and repeated at 0.3 K with *a*-axis–oriented domains initially prepared via out-of-plane FSR. The resulting MFM images reveal an approximately perpendicular relationship between the magnetic domain orientation $\theta_{MD}$ and $\theta_B$ (Fig. S8), consistent with the Bloch-type domain wall structure typically existed in magnetic bulk crystals.

The extracted peak positions from the fitted $I_{FFT}(\theta_{FFT})$ curve after in-plane FSR at 4.2 K reveal a stronger tendency toward *a*-axis orientation: FSR conducted along *a*/*b*-axis tend to induce magnetic domains with pure perpendicular component, while magnetic field along $\theta_B$ = 25°, 45°, 65° cause coexistence of such perpendicular component and a significant *b*-axis orientated magnetic domain component (Fig 3c). The lasting *b*-axis orientation component of $EuTi_3Bi_4$ indicates a consistent tendency for magnetic domains to align along *b*-axis at 4.2 K, revealing lower magnetic energy while extending along *b*-axis.

To further analyze the relation between in-plane magnetic-field regulation and magnetic domain orientation, the field regulation factor $\eta$ is defined as the absolute difference between $\theta_B$ and $\theta_P$ (the position of the dominant peak in the $I_{FFT}$ curve): $\eta = |\theta_P - \theta_B|$. Taking into account the tendency of stripe domains to align perpendicularly to the in-plane magnetic field, along with the 90° rotation between the stripe orientation in the MFM image and the corresponding features in the FFT image ($|\theta_P - \theta_{MD}|$ = 90°), $\eta$ serves as a sensor of the regulation efficiency of an in-plane magnetic field applied in a given direction. A smaller value of $\eta$ indicates higher regulation efficiency and lower energy for magnetic domains extending perpendicular to $\theta_B$. The extracted $\eta(\theta_B)$ curves further reveal the variation of anisotropic in-plane magnetic domain energy at 0.3 K and 4.2 K (Fig 3d). At both temperatures, $\eta(\theta_B)$ reaches minimum values at $\theta_B$ = 0° and $\theta_B$ = 90°, indicating that magnetic domains aligned along the crystallographic *a*-axis and *b*-axis correspond to local energy minima. More importantly, a notable difference emerges between the two temperatures. At 0.3 K, $\eta$(0°) is slightly lower than $\eta$(90°), suggesting that magnetic domains aligned along the *a*-axis are energetically favored over those along *b*-axis. In contrast, the in-plane MAS mapping reverses at 4.2 K, where the energy of *b*-axis-oriented domains becomes lower.

To understand the origin of the magnetic anisotropy in $EuTi_3Bi_4$ and the mechanism of its temperature-driven texture switching, we performed DFT calculations. Several candidate spin configurations were compared (Fig. S9), and the ferromagnetic state was found to have the lowest energy, lying at least 0.54 meV/Eu below the competing antiferromagnetic configurations (Fig. S10a and Table S1), consistent with the experimentally observed ferromagnetic order [31, 40]. Based on this ferromagnetic ground state, we calculated the magnetic anisotropy energy. The magnetic anisotropy energy (MAE) landscape exhibits a global minimum along the *c*-axis, identifying it as the magnetic easy axis. Within the *ab* plane, the *a*-axis is the second most stable direction and the *b*-axis direction is most unfavored in energy, with $E_a - E_c = 2.22$ meV/Eu and $E_b - E_c = 2.68$ meV/Eu. These qualitative conclusions—the ferromagnetic ground state, the *c*-axis easy axis, and the ordering $E_c < E_a < E_b$—are robust across the effective on-site Coulomb interaction parameter $U_{eff} = 0$–4 eV (Figs. S10a and S10b, Table S2). For the discussion that follows, we adopt $U_{eff} = 1$ eV, which gives the best agreement with the experimental coercive field (Fig. S8).

The magnetic anisotropy mainly contains contributions from three distinct mechanisms [1]: magnetostatic shape anisotropy originating from dipole-dipole interactions, single-ion anisotropy from spin-orbit coupling, and anisotropic exchange. For bulk $EuTi_3Bi_4$, the shape anisotropy is relatively weak compared with the single-ion and exchange contributions, and its variation over the low-temperature range of interest is negligible compared with the observed texture switching. We therefore focus on the latter two contributions by mapping the calculated magnetic energies onto a spin Hamiltonian:

$$H=-\frac{1}{2}\sum_{ij}\mathbf{S}_i^T\mathbf{J}_{ij}\mathbf{S}_j-\sum_{i}\mathbf{S}_i^T\mathbf{A}_i\mathbf{S}_i\,,$$

where $A_i$ ($i$ = a, b, c) are the single-ion anisotropy constants along the three crystallographic axes, and $J_{ij}$ is the exchange interaction matrix between nearest neighbor sites, whose diagonal elements $J_{aa}$, $J_{bb}$, $J_{bb}$ describe the anisotropic exchange along each axis. To quantify the in-plane anisotropy, we define $\Delta A_{ab} = (A_a - A_b)S^2$ and $\Delta J_{ab} = (J_{aa} - J_{bb})S^2$, where $S$=7/2 is the spin magnitude of $Eu^{2+}$. At $U_{eff}$=1 eV, the extracted values are $\Delta A_{ab}$=0.55 meV/Eu and $\Delta J_{ab}$=-0.20 meV/Eu. Their opposite signs show that the single-ion anisotropy favors the *a*-oriented spin component, whereas the anisotropic exchange favors the *b*-oriented component. In comparison, the single-ion anisotropy along *c* is substantially stronger, with $(A_c\text{-}A_a)S^2$=2.28 meV/Eu and $(A_c\text{-}A_b)S^2$=2.83 meV/Eu. This dominant *c*-axis contribution stabilizes the magnetic easy axis along *c*, preventing the competing in-plane anisotropies from reorienting it. These opposite signs hold throughout the investigated U value range (Fig. 3g, Fig. S10c-e and Table S3, S4). The energy difference between the two in-plane hard axes, set by the competition between A and J, determines the shape of the elliptical trajectory traced by the precessing Eu moments at finite temperature: a larger a-axis component of the trajectory favors a-oriented domain textures, while a larger b-axis component favors *b*-oriented ones (Fig. 3h). At low temperature, *A* dominates, giving an a-axis preference, but since $A$ and $J$ are comparable in magnitude, a modest thermal or magnetic stimulus can tip this balance. The single-ion anisotropy $A$ is sensitive to the electronic states near the Fermi level [32,41], whereas the exchange anisotropy $J$ is controlled by intersite orbital hybridization and is less affected by thermal effect [45]. Consequently, the effective anisotropy contribution associated with A scales approximately as $M^3$, whereas the contribution associated with J approaches an $M^2$ dependence as temperature increases and exchange anisotropy becomes dominant [46]. At the switching temperature $T_s$, J overtakes A, and the preferred domain texture reorients from the a-axis to the *b*-axis.

We also considered alternative explanations. The calculated phonon dispersion contains no imaginary modes, ruling out a soft-phonon-driven CDW transition (Fig. S11) [47]. Quasi-harmonic calculations [48] give a relative volume expansion of only $10^{-4}$ from 0 to 15 K, too small to explain the texture switching (Fig. S12). Constrained-angle calculations confirm that increasing the canting angle alone does not reverse the in-plane preference (Fig. S13), excluding the possibility that the switching merely reflects a temperature-driven change in the canting amplitude. The switching is therefore a genuine consequence of the temperature-dependent competition between single-ion and exchange anisotropies, through which a small change in their relative strength reorients the in-plane domain texture while the *c*-axis easy magnetization remains intact.

## V. Magnetic domain anisotropy switch in $EuTi_3Bi_4$ flake

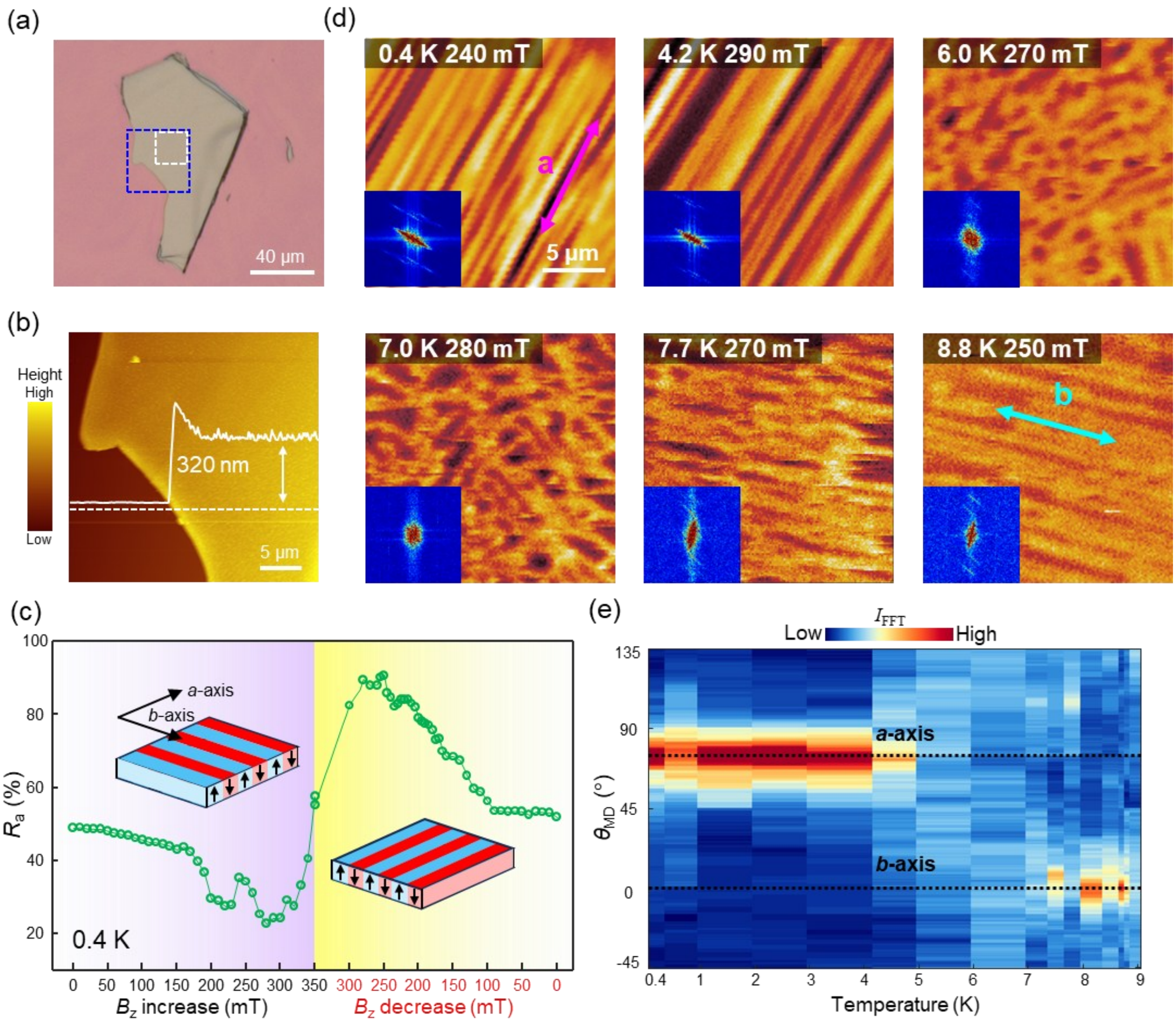

**Fig.4. Temperature-dependent domain orientation switch of $EuTi_3Bi_4$ flake.** (a) Optical image of $EuTi_3Bi_4$ flake (b) Topography image of the $EuTi_3Bi_4$ flake measured by AFM. (c) $B_z$–dependent *a*-axis ratio component $R_a$ curve at 0.3 K in a $EuTi_3Bi_4$ flake. (d) MFM images of magnetic domains in $EuTi_3Bi_4$ flake during FSR at different temperatures, measured in the white dashed box in (a). Insets show corresponding FFT images. (e) Temperature-dependent magnetic phase diagram revealing significant magnetic domain orientation switch at 6~7 K.

To further manipulate the MDS effect of $EuTi_3Bi_4$. A 320 nm thick $EuTi_3Bi_4$ flake is obtained via an Au-assisted mechanical exfoliation method (Fig 4a-b). Out-of-plane FSR operation of the flake reveals a consistent saturation field of $\boldsymbol{B}_z$ = 350 mT, and reproduces MDS phenomenon at 0.3 K (Fig. S14), underscoring advantage of the layered structure. Meanwhile, the composite magnetic domain structure is degraded into a pure stripe phase in the $EuTi_3Bi_4$ flake. The extracted $R_a$ curve represents a significant magnetic domain orientation switching during the FSR process (Fig. 4c), closely resembling that of bulk $EuTi_3Bi_4$. Notably, the MFM signals of magnetic domains are comparatively weak ($R_a \approx 50\%$ at low $\boldsymbol{B}_z$) in the $EuTi_3Bi_4$ flake. At temperatures above 4.2 K, magnetic domains emerge only within a narrow field range during the FSR process and disappear at $\boldsymbol{B}_z = 0$ T (Fig S15).

At $T < 6$ K, FSR induced stripe domains align along *a*-axis, which switch to *b*-axis at $T > 7$ K. In the intermediate range of 6 K < $T$ < 7 K, disordered magnetic domains with competing *a*/*b* axis components are observed (Fig 4d). The corresponding FFT images and extracted phase diagram (Fig 4e) confirm the presence of MDS behavior in the $EuTi_3Bi_4$ flake, with a significantly increased switching temperature of $T_s \approx 7$ K. It is worth noting that the *a*-axis and *b*-axis labelled in Fig 4c are separated by only 75°, likely due to deformation of the flake introduced during exfoliation. Complete MFM and FFT images of temperature-dependent magnetic domain textures in $EuTi_3Bi_4$ flake after *c*-axis FSR are listed in Fig S16.

The observed increase of $T_s$ in $EuTi_3Bi_4$ flake is likely attributed to introduced strain during exfoliation process [44]. This additional strain shortens the Eu–Eu distance between the zigzag chains, thereby requiring a higher temperature to expand this distance and enable the reorientation of magnetic domains from the *a*-axis to the *b*-axis. Sample thickness may also contribute to the marked increase in the critical temperature. The persistence of the MDS effect in exfoliated $EuTi_3Bi_4$ highlights its potential for future applications in low-dimensional spintronic devices.

## VI. Conclusion

We reported a novel MDS phenomenon in $EuTi_3Bi_4$ crystal. The energetically preferred magnetic domain orientation switches between *a*-axis and *b*-axis of the lattice under *c*-axis magnetic field and temperature modulation. Magnetization measurements and DFT calculations confirm a robust *c*-axis easy magnetization, indicating that the observed *a*/*b* switching corresponds to an in-plane reorientation of magnetic domain textures rather than an easy-axis reorientation. The calculated single-ion anisotropy tensor $A$ and nearest-neighbor anisotropic exchange tensor $J$ exhibit opposite in-plane preferences, further suggesting a possible competition between the *a*-favoring single-ion anisotropy and the *b*-favoring exchange anisotropy and providing a qualitative microscopic picture for the temperature-dependent texture switching. We further confirm the persistence of the MDS in an exfoliated $EuTi_3Bi_4$ flake, accompanied by a significant increase in the switching temperature $T_s$, highlighting its high tunability and promising potential for spintronic applications. The capability to flexibly modulate magnetic domain orientations through temperature or magnetic fields, without incurring the high energy cost of reorienting the easy axis, offers a promising route for low-cost control of magnetic and electronic properties in exfoliated van der Waals magnets and magnetic heterostructures.

## Acknowledgements

We gratefully acknowledge the financial support from the National Natural Science Foundation of China (12522408, 62488201, 12374199, 92477205, 52461160327 and 92463307), the Ministry of Science and Technology (MOST) of China (2022YFA1204100, 2023YFA1406500 and 2025YFA1212800), and the Beijing Nova Program (Nos. 20240484651). Calculations were performed at the Physics Lab of High-Performance Computing (PLHPC) and the Public Computing Cloud (PCC) of Renmin University of China.